\documentstyle[a4,12pt,axodraw,amssymb]{article}
\let\vp=\varphi 
 
\let\al=\alpha

\let\Ga=\Gamma 
\let\de=\delta 
\let\De=\Delta
\let\ep=\varepsilon 
 
\let\ka=\kappa 
\let\la=\lambda
\let\La=\Lambda 
 
\let\si=\sigma

\let\Om=\Omega
 
\renewcommand{\(}{\left(} \renewcommand{\)}{\right)}
\renewcommand{\[}{\left[} \renewcommand{\]}{\right]}
\renewcommand{\thefootnote}{\fnsymbol{footnote}} 
\def\vev#1{\langle #1\rangle} 
\newcommand{\beq}{\begin{equation}}
\newcommand{\eeq}{\end{equation}} 
\newcommand{\bea}{\begin{eqnarray}}
\newcommand{\eea}{\end{eqnarray}} 

\newcommand{\gev}{\;{\rm GeV}}
\newcommand{\tev}{\;{\rm TeV}} 
\newcommand{\mgut}{M_{\rm GUT}}
\newcommand{\mg}{m_{3/2}}

\newcommand{\Snu}{\widetilde{N}}

\newcommand{\M}{M_*}
\newcommand{\ie}{{\it{i.e. }}}
\newcommand{\eg}{{\it{e.g. }}}
\newcommand{\oo}{${\cal O}$}

\begin{document}
\thispagestyle{empty}
\vspace*{.5cm}
\noindent
\hfill {\tt hep-ph/0208003}\\
\vspace*{1.5cm}
\begin{center}
{\Large\bf Leptogenesis at Low Scale}\\[1cm] {\large Lotfi
Boubekeur}\footnote[5]{E-mail: {\tt lotfi@he.sissa.it}}\\[.2cm]

{\it  SISSA-ISAS,  Via Beirut 4, I-34013 Trieste, Italy.\\ and\\ INFN
Sezione di Trieste, Trieste, Italy. }
\end{center}
\vskip 1 cm

\date{\today}

\begin{abstract}

A typical problem of the leptogenesis scenario is the mismatch 
between the maximum reheat temperature implied by gravitino
overproduction bound and the minimum temperature required to create
thermally the lightest right-handed neutrino. We explore the
possibility of baryogenesis via leptogenesis in the presence of low
scale mass right-handed neutrino. In such a scenario, right-handed
neutrinos are created thermally at low reheat temperatures 
without relying on non-perturbative production mechanisms.  
We focus on two specific realizations of the scenario, namely the 
out-of-equilibrium decay of right-handed neutrinos (Fukugita-Yanagida) 
and the leptogenesis via the $LH_u$ flat direction (Affleck-Dine). 
We find that in general, the two scenarios are able to produce 
the required baryon excess for a reasonable amount of CP violation.

\end{abstract}

\newpage
\setcounter{page}{1}
\renewcommand{\thefootnote}{\arabic{footnote}}

\section{Introduction} 
Recent experimental results gave overwhelming evidence that neutrinos
have small but non-vanishing masses \cite{SuperK}.  In the standard
model (SM), neutrinos are exactly massless, hence the  explanation of
neutrino experiments requires Physics beyond the SM.   Furthermore,
neutrino masses appear to be very small with respect to the  other
fermions ones. If neutrinos are Majorana particles, it is possible 
to accommodate small neutrinos masses in the SM by introducing the 
lepton number violating effective operator \cite{weinberg} $ O_{\rm eff}=
\al_{ij}\ell^T_i\tau_2 \vec{\tau} \ell_j H^T \tau_2 \vec{\tau} H /M $,
where $\ell_i$ and $H$ are the lepton and the Higgs doublet
respectively.  Here $M$ is the scale where ``new Physics'' is expected
to occur, is usually  taken as the Planck or the GUT scale. In the
former case, the presence of  this lepton number violating operator is
motivated by the common belief  that gravity does not respect any
global quantum number  \cite{akhmedov,barbieri}, or at least this is
what happens for example in  black holes and wormholes --no hair
theorems.  In the latter case  ($M=\mgut$), the effective operator
arises via the see-saw mechanism  \cite{seesaw}, when integrating-out
the heavy right-handed neutrinos (RHNs hereafter).
\vskip 4pt
On the other hand, our Universe appears to be constituted exclusively
of baryons. In order not to spoil the Big Bang Nucleosynthesis (BBN)
successful  predictions of the observed light elements abundances
\cite{olive}, a small baryon excess have to be present. The required
value is quantified by the baryon-to-entropy ratio and is given by
$Y_B\equiv n_B / s=(7.2 \pm 0.4)\times 10^{-11}$~\cite{pdg}. 
To accomplish successfully their task, baryogenesis
scenarios~\cite{review} have to satisfy three essential conditions
\cite{sakharov}, namely:   {\em (i.) Baryon number violation}, {\em
(ii.) $C$ and $CP$ violation}, and {\em(iii.) Departure from thermal
equilibrium}. One particularly appealing scenario is the leptogenesis
scenario, where lepton number,  produced either by the
out-of-equilibrium decay of heavy RHN's \cite{FY}  or by the decay of
a scalar condensate carrying non-zero lepton number  \cite{AD,drt}, is
reprocessed to a baryon asymmetry via the sphalerons
interactions. Given the experimental evidence that lepton number is
violated  in neutrino oscillation and the fact that proton decay have
not been observed yet, the present experimental situation seems to
favor this scenario over the other existing baryogenesis scenarios.
\vskip 4pt
A generic problem of thermal leptogenesis scenarios is the mismatch
between  the maximum reheat temperature implied by gravitino
overproduction and the  minimum temperature required to thermally
create heavy RHNs  $T_{\rm RH}\gtrsim
10^{10}\gev$. To reconcile these two facts, non-thermal creation of
RHNs in a low reheat temperature plasma were considered. These
mechanisms,  however, involve non-perturbative dynamics and are in
general sensitive to  inflation models. Furthermore, they lead to even
more stringent bounds on  the reheat temperature, due to the non
thermal production of moduli and  gravitinos
\cite{nonthermal,Giudice:1999am}.

\vskip 4pt
The aim of this paper is to address this issue in a different
perspective.  We will consider the situation where the reheat
temperature is low (may  be as low as the TeV) and we will only
consider thermal production of RHNs. This will naturally lead us to
consider a class of see-saw models  (that we will subsequently call
low-scale see-saw models), where RHNs have  TeV masses instead of the
conventional unification scale.
\vskip 4pt
The paper is organized as follows. In section \ref{sec:grav-probl-vs},
we give our main motivation for the scenario. In section
\ref{sec:therm-lept}, we study leptogenesis through the
out-of-equilibrium decay of low scale RHNs. In section
\ref{sec:ad}, we turn to the Affleck-Dine scenario. Finally, in
section \ref{sec:concl}, we summarize our conclusions.

\section{The gravitino problem vs. thermal leptogenesis}
\label{sec:grav-probl-vs}
As any unwanted relic, gravitinos represents a potential danger for
the  thermal history of the Universe. Gravitinos are created
predominantly via  $2\to2$ inelastic scatterings of gluons and gluinos
quantas. Their relic density and contribution 
to the energy density are given by \cite{production} 
\bea
\label{grav}
&Y_{3/2}&=1.1 \times 10^{-10}\( T_{\rm RH}\over 10^{10}\gev \) \(100
\gev\over \mg\)^2 \( m_{\tilde{g}}\over 1\tev\)^2\,, \\
\label{Omega}
&\Om_{3/2}\, h^2 &=0.21 \( T_{\rm RH}\over 10^{10}\gev\)\(100\gev\over\mg\)
\(m_{\tilde{g}}\over1 \tev\)^2\,,
\eea

where $m_{\tilde{g}}$ denotes the gluino mass. The requirement that, 
if unstable, their late decay do not disrupt the successful BBN
predictions, and if stable,  their energy density do not overclose 
the Universe, put tight constraints  on their relic abundance. It has 
been noted that if $\mg>$10 TeV or  $\mg<$ keV, then there is no
gravitino problem \cite{Weinberg:zq,Pagels:ke}.  These requirements
can be relaxed if there is a period of inflation and the  constraints
apply only on post-inflation abundances. From the expression
(\ref{grav}), one sees that the gravitino abundance scales linearly
with the reheat temperature, therefore the bound on $Y_{3/2}$
translates onto the following bound on the maximum allowed reheat 
temperature $T_{\rm RH}$ \cite{gravitino}  
\beq 
T_{\rm RH}\lesssim (10^6-10^9)\gev \mbox{ for }
\mg=100\gev-1\tev\,.  
\eeq 
There exists however more stringent bounds
on  $T_{\rm RH}$ from non-thermal production. For generic
supersymmetric inflation models, the bound can be as tight as
\cite{Giudice:1999am} $T_{\rm RH}\lesssim 10^5 (V^{1/4}/10^{15}\gev)$,
where $V^{1/4}$ is the height of the inflationary potential.
\vskip 4pt
Let us now see the constraints on the reheat temperature coming from
leptogenesis. In the original see-saw model \cite{seesaw} the mass
scale of  RHNs is typically of \oo $(10^{10}-10^{15}) \gev$. In
addition, the bound on  the CP parameter \cite{sacha} for hierarchical
RHN's in thermal leptogenesis  implies a lower bound on the mass of
the lightest RHN  $M_{N_1}\gtrsim 10^{10}$ GeV.  Consequently, if the
thermal leptogenesis scenario is truly  {\it {the mechanism}}
responsible for the the generation of the Baryon  Asymmetry of the
Universe (BAU), RHNs of this mass have to be produced  after
inflation. This means a high reheat temperature, at least as high  as
the mass of the lightest RHN, \ie $10^{10} \gev$, potentially
conflicting  with the gravitino bound discussed above.
\vskip 4pt
A possible way out to get around this problem is to produce RHNs
non-thermally, that is during  an efficient preheating phase 
\cite{gprt}. Non-thermal production,
however, can lead in some cases to even more stringent  bounds on the
reheat temperature. Indeed, for typical hybrid inflation  models, the
upper bound on the reheat temperature can be as low as 1 TeV
\cite{nonthermal}.

\vskip 4pt
From the above discussion, it is clear that any compelling solution to
this  problem will, in one way or another, involve low reheat
temperatures. After  all, we dont know the thermal history of our
Universe before BBN. All we know  experimentally is that $T_{\rm
RH}\ge T_{\rm BBN}\sim$ MeV. In this paper,  we will consider a rather
exotic solution to this problem, namely the case for leptogenesis
when RHNs have a low scale mass. The first benefit of such an
approach is that RHNs can be produced thermally with a low reheat
temperature $T_{\rm RH}\sim$ \oo(TeV), avoiding  thus the creation of
dangerous relics, like heavy GUT monopoles, and more importantly 
suppressing the creation of gravitinos.  On theoretical grounds, 
nothing forbids the mass of RHNs to be of \oo (TeV).  In fact this 
situation is encountered in many cases  (See for \eg
\cite{Arkani,Borzumati,dudas,nasri}). This is also a  typical
situation that arises in models where the fundamental scale (the GUT
scale and/or the quantum gravity scale) is of \oo (TeV). In this case,
the  Yukawa couplings of RHNs have to be much smaller to produce
phenomenologically acceptable light neutrino masses. Such a
fine-tuning is stable under radiative corrections because that Yukawa
couplings are self renormalizable and is protected by
supersymmetry. There remains the question of how such suppressed
Yukawa couplings can arise in a concrete model. This can be achieved
for example by the mean of some $R$-symmetry  that forbids the bare
Yukawa coupling between the left and the right-handed  neutrinos. As a
result the leading Yukawa couplings will be suppressed by  powers of a
heavy scale \cite{Arkani,Borzumati}. The Yukawa suppression can be
obtained upon integrating-out some heavy field as well \cite{FN}.

\section{Thermal Leptogenesis with TeV scale RHNs} 
\label{sec:therm-lept}
We begin by reviewing the basics of the out-of-equilibrium decay
leptogenesis scenario. Consider the Minimal Supersymmetric
Standard Model extended by three RHNs, one for each generation.  The
interactions of the RHNs are given by the following superpotential
\beq 
W_N=Y_{ij} L_i H_u N_j + {1 \over 2} M_i N^2_i  
\eeq
After integrating-out the RHN and electroweak symmetry breaking, the light
neutrinos mass matrix will be given by the familiar see-saw formula
\beq
m_\nu = - Y^T M^{-1} Y \vev{H_u}^2\,. 
\eeq
In this scenario, the RHNs must decay
out-of-equilibrium. A measure  of the departure from thermal
equilibrium is given by the parameter $K$  defined as  
\beq 
K \equiv
\frac{\Gamma_{N}}{2\,H} {\Big\vert_{T=M_N}}\,,  
\eeq
where $\Ga_N$ is
the decay rate of RHNs and $H$ is the expansion rate of the
Universe. The decay is out-of-equilibrium when $K \lesssim 1$. The final
baryon  asymmetry reprocessed by sphalerons is given by
\cite{sphalerons}
\beq   
Y_B\equiv\frac{n_B}{s}=\(\frac{8n_g+4 n_H}{22 n_g+ 13
n_H}\)\frac{n_L}{s}\,,   
\eeq
where $n_g$ and $n_H$ counts the number of fermion generations and
Higgses respectively. 

The lepton asymmetry produced by
the CP-violating out-of-equilibrium decay of the RHNs can be computed
using 
\beq  
{n_L \over s }=\ka\, {\ep \over g_*}\,,  
\eeq 
where $g_*$ is the effective degrees of freedom and $\ka$ is the dilution 
factor, computed by integrating the relevant set of Boltzmann equations
\cite{boltzmann,Plumacher:1996kc}.  The parameter $\ep$ characterizing
CP violation in the RHNs decay, can be defined for each RHN separately
as \cite{CP}
\bea  
\ep_i &\equiv& {\sum_j \Ga(N_i\to \ell_j h_u)- \sum_j \Ga(N_i\to
\bar{\ell}_j \bar{h}_u) \over \sum_j \Ga(N_i\to \ell_j h_u)+ \sum_j
\Ga(N_i\to \bar{\ell}_j \bar{h}_u)} \nonumber \\  &=& -{1\over 8 \pi}{
1 \over \(Y Y^\dagger \)_{ii}} \sum_{k\neq i} \mbox{Im}
\[ \{ (Y Y^\dagger)_{ik}\}^2 \] 
 \[ F_V \(M^2_k\over M^2_i\) + F_S \(M^2_k\over M^2_i\)  \]
\label{ep}
\eea
where $F_V$ and $F_S$ are the contributions of the vertex and
self-energy  respectively. They are given by  
\beq
\label{f}
F_V(x)=\sqrt{x} \ln \( 1 + { 1 \over x  }\), \;\;\;  F_S(x)= {2
\sqrt{x} \over x-1}  
\eeq
Now, applying the above formulae to TeV mass RHNs, one immediately
sees that,  due to the smallness of the Yukawa couplings, the decay of
RHNs is  automatically out-of-equilibrium. In addition to the decay
processes, there  can be other competing processes that might bring
the RHNs to thermal equilibrium, depleting any pre-existing lepton
number.  These processes have to be out-of-equilibrium too, \ie
$\Ga\simeq\vev{n\si v}\ll H$. The first such process is the $\De L=2$
scattering $\ell h_u \leftrightarrow \bar{\ell} \bar{h_u}$, via both
$s$ and $t$ channel. Other competing processes may involve the
$t$-($s$)quark, such as $N t(\bar{b}) \leftrightarrow \ell
b(\bar{t})$. It turns out that due to the Yukawa coupling suppression
all these processes are out-of-equilibrium.  Finally, it has been
noted \cite{sarkar} that the process $W^{\pm}W^{\pm} \to
\ell^{\pm}\ell^{\pm}$, mediated by virtual left-handed neutrinos can
lead to stringent constraints on their masses.  In our case, it leads
to a very mild constraint.
\vskip 4pt
So far for the out-of-equilibrium conditions, now we concentrate on
the CP violation parameter $\ep$. As we have seen previously, due to
the smallness of the Yukawa couplings, it is very easy to satisfy the
out-of-equilibrium condition, however the resulting CP violation
parameter $\ep$ is too small. This is due to the fact that the decay
rate $\Ga$ and the CP-parameter $\ep$ are both proportional to the
same Yukawa couplings combination. From Eqts (\ref{ep}, \ref{f}), one
sees that the two contributions to the CP parameter $\ep$ are sensible
to two completely different patters of RHNs masses. While the vertex
contribution $F_V$ is enhanced for large hierarchies, the self-energy
contribution $F_S$ is so when RHNs are (quasi-)degenerate
\footnote{\label{foot}  For perturbation theory to hold, the mass
splitting $\de M_{ik}=|M_i-M_k|$ must satisfy $\de M_{ik}\gg  \Ga$,
where $\Ga$ is the decay rate of RHNs, otherwise one can no more trust
the perturbative calculation based on Eqts (\ref{ep},\ref{f}) and  one
have to rely on a resummation approach \cite{pilaftsis}. In the limit
of exact degeneracy, the CP parameter vanishes.}. In order to enhance
the value of $\ep$, one have to exploit the properties of the two
functions $F_V$ and $F_S$. In the next subsection, we will consider
the case where RHNs are nearly degenerate
\cite{hambye,pilaftsis}. There  exist however another possibility,
related to the fact that RHNs masses  and soft SUSY breaking $A$-terms
are of the same order \ie \oo(TeV).

\subsection{Leptogenesis with quasi-degenerate TeV scale RHNs}
\label{sec:deg}
Consider a model where two out of the three RHNs are quasi degenerate,
that is $N_1$, $N_2$ and $N_3$ have masses $M_1,\,M_2 \sim$ \oo(TeV)
$\ll M_3$ respectively.  The mass splitting $\de M_{12} \equiv |M_2-M_1|=\de
\cdot M_0$, where $M_0\sim$~TeV).  Due to their suppressed Yukawa's,
RHNs will be long-lived enough to eventually  dominate the Universe
before decaying. The condition for RHNs  dominance can be written
$\Ga_N\ll \Ga_\vp$, where $\Ga_\vp$ is the decay  rate of the inflaton
\cite{HMY}. While, RHNs can hardly dominate the energy  density of the
Universe because of Pauli blocking, this can happen more  easily for
their scalar partners the RH sneutrinos. Moreover, due to quantum  de
Sitter quantum fluctuations \cite{Linde:uu} and for  $H_{\rm inf}\gg
M_N \sim$ \oo(TeV), RH sneutrinos become coherent over  super-horizon
scales and can be considered as classical fields with the  constant
value (vev) $\vev{\Snu^2}=3 H^4_{\rm inf}/8\pi^2 M_N^2$. Therefore  if
the RH sneutrinos scalar potential is just given by the mass term,
they  are likely to dominate quickly the energy density of the
Universe. Given  the above discussion, one can compute the lepton
asymmetry produced during  the decay of $N_2$ using  
\beq 
{n_L \over s}={3 \over 4} {T_{N_2}\over M_2} \ep_2 \,,  
\eeq 
where $T_{N_2}$ is
the decay temperature of $N_2$'s computed by equating the energy
density of RHNs with the energy density of the Universe when 
$H \sim \Ga_2$.  Since $N_1$ and $N_2$ are quasi-degenerate, 
we can safely ignore the vertex contribution to  the CP parameter 
($F_S \gg F_V$). Using Eqts (\ref{ep})
and (\ref{f}), we can  compute the total CP parameter
$\ep\simeq\ep_1+\ep_2$, giving  
\beq 
\ep\simeq {1\over 8 \pi}
\sum_{i=1,2} {1 \over (YY^\dagger)_{ii}}\, \mbox{Im}
\[ \{(YY^\dagger)_{12}\}^2 \]
{1 \over \de} 
\eeq 
A rough estimate of the required degeneracy gives
$\de\sim$ \oo $(10^{-6}-10^{-7})$, and perturbativity is clearly
satisfied (See footnote  \ref{foot} on page \pageref{foot}).  Such a
degeneracy could be ascribed for example to a flavor symmetry, the
parameter $\de$ would then characterize its breaking.  In the simplest
case, the flavor group $G_f$ is taken as a $Z_2$ and the RHNs have
different parity $Z_2$ assignments, \ie $N_1\sim$ odd (even) and
$N_2\sim$ even (odd) under $Z_2$.  The flavor symmetry is broken by
the vev of the odd field $\psi$.  Restricting to the 12 block, the
resulting mass matrix for the RHNs is 
\beq 
M_R \sim M_0 \(
\begin{array}{cc}
1 & \de /2 \\ \de/2 & 1
\end{array}
\) 
\eeq 
with $\de/2 \equiv \vev{\psi}/ \La$. The diagonalization of
the mass matrix  yields two quasi-degenerate RHNs with a
mass-splitting  $\de M_0$.

\subsection{Leptogenesis from soft SUSY breaking $A$-terms}
\label{sec:soft}
In the traditional leptogenesis scenario, the contributions of the
soft  SUSY breaking $A$-terms to the CP parameter $\ep$ are usually
neglected.  Indeed, SUSY breaking will induce the following $A$-terms
\beq
\label{soft}
{\cal L}_{\rm soft}= A_{ij} \mg 
\widetilde{L_i}  \Snu_j H_u + {\rm h. c.}\,.
\eeq 
Let us consider the following vertex diagrams, where in the tri-scalar
vertex we put the $A$-term contribution from Eqt (\ref{soft}) instead
of the standard SUSY one. \\

\begin{center}


\begin{picture}(300,90)(0,0)
\ArrowLine(0,45)(40,45) \ArrowLine(40,45)(65,70)
\DashLine(40,45)(65,20){4} \DashLine(65,70)(65,20){4}
\DashArrowLine(65,20)(105,20){4} \ArrowLine(65,70)(105,70)
\Text(20,55)[]{$N_i$} \Text(45,65)[]{$\widetilde{H}$}
\Text(45,30)[]{$\widetilde{L}_k$} \Text(75,45)[]{$\widetilde{N}_l$}
\Text(85,80)[]{$L_j$} \Text(85,30)[]{$H$} \Text(65,0)[b]{$(a)$}


\DashArrowLine(150,45)(190,45){4} \Text(170,55)[]{$N_i$}
\DashLine(190,45)(215,70){4} \Text(195,65)[]{$\widetilde{H}$}
\DashLine(215,70)(215,20){4} \Text(225,45)[]{$\widetilde{N}_l$}
\DashArrowLine(215,20)(255,20){4} \Text(195,65)[]{$\widetilde{H}$}
\DashArrowLine(215,70)(255,70){4} \DashLine(190,45)(215,20){4}
\Text(195,30)[]{$\widetilde{L}_k$} \Text(235,30)[]{$H$}
\Text(235,80)[]{$\widetilde{L_j}$} \Text(215,0)[]{$(b)$}

\end{picture}
\vskip 1 cm
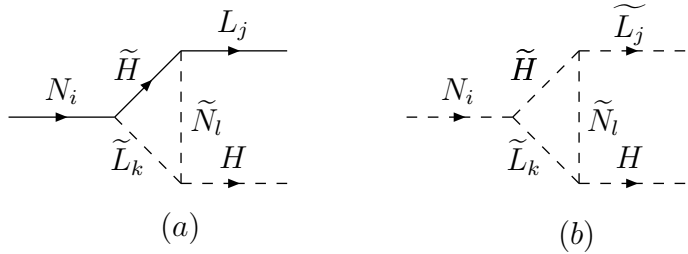
{\small  {\bf Figure 1:} SUSY breaking $A$-term contributions to the
CP parameter $\ep$.}
\end{center}

Estimating the contribution of the SUSY soft breaking $A$-terms to
$\ep$ and comparing it to the standard SUSY one for each of the two
considered diagrams, we obtain  
\bea
\label{epsoft}
{\ep^{\rm soft}_{(a)}\over \ep^{\rm SUSY}}&\sim& |A| {\mg \over M_i} \sin
\de_{\rm soft}\,,\\
{\ep^{\rm soft}_{(b)}\over \ep^{\rm SUSY}} &\sim& |A|^4 \({\mg \over M_i}\)^4 \sin
\de_{\rm soft}
\eea 
where $\de_{\rm soft}$ is an effective soft CP
phase. From (\ref{epsoft}),  we see that in the conventional
leptogenesis scenario, where the mass of the  lightest RHN is $M_1
\simeq 10^{10}$ GeV, $\ep_{\rm soft}$ is suppressed  with respect
to $\ep^{\rm SUSY}$ at least by a factor of $10^{-7}$. However, in our 
scenario, where  $M_i \sim \mg$, the CP asymmetry parameter  $\ep_{\rm
soft}$ is no more suppressed. It can even dominate the over  the SUSY
contribution depending on the value of the soft parameters. This means
that CP violation may completely originate from the soft SUSY breaking
sector, like in the Affleck-Dine case. However, besides enhancing the
amount of CP violation, the soft SUSY breaking interactions could
bring the RHNs decay at equilibrium, erasing considerably the
produced lepton number. A more accurate analysis, requiring the
integration of Boltzmann equations, is necessary to reach a firm 
conclusion.
\vskip 4 pt

Finally, it is worth noticing from Eqt. (\ref{Omega}) that gravitinos 
could no more constitute a sizable amount of dark matter in our
scenario. Indeed, $\Om_{3/2}\,h^2=0.01-1$, requires the gravitino to
be lighter and/or the gluinos masses to be heavier.

\section{Affleck-Dine leptogenesis with TeV scale RHNs}
\label{sec:ad}
Now, we turn to investigate the Affleck-Dine mechanism \cite{AD,drt}
in the  presence of TeV scale RHNs. Consider the $LH_u$ MSSM flat
direction given by \footnote{The factor $\sqrt{2}$ is necessary to 
have a canonical kinetic term for $\vp$ (The Kahler potential  is 
$K= H_u H_u^\dagger+ L L^\dagger= \vp\vp^\dagger$).}  
\beq
\label{lh}
L_i ={1 \over \sqrt{2}}\(
\begin{array}{c}
\vp \\  0
\end{array}\), \;\;\;
H_u ={1 \over \sqrt{2}}\(
\begin{array}{c}
0 \\ \vp
\end{array}\)
\eeq
This flat direction is lifted by the non-renormalizable operator
$W_{\rm NR} =\la (L\,H_u)^2 /M = \la  \vp^4/4 M$. This operator can be
generated via the see-saw mechanism when integrating-out  the heavy
RHNs. The evolution of the scalar condensate $\vp$ in the expanding
background is dictated by the classical equation of motion
\beq
\ddot{\vp}+3 H \dot{\vp} + {\partial V(\vp) \over \partial \vp^*}=0
\eeq
where $V(\vp)$ is the full potential, including the soft masses, the Hubble 
induced masses and the $A$-terms (both from SUSY breaking and the Hubble 
induced ones)\footnote{Here, we are simply ignoring thermal effects
\cite{thermal}.}.
\bea
V(\vp)&=&(\mg^2 - c_H H^2)|\vp|^2 +  a_H H {\vp^4 \over 4 M \nonumber}\\ 
&& + a_m \mg {\vp^4 \over 4 M} + \mbox{h.c.}+ {|\la|^2 \over M^2} |\vp|^6.
\label{V}
\eea
The constants $c_H$ and $a_H$ depend on the detailed structure of the Kahler 
potential. In particular the sign of $c_H$ is crucial for the validity of the 
AD scenario. We assume throughout the paper that it is positive ($c_H>0$). 
The evolution of the scalar condensate follows three phases. During 
inflation, when $H \gg \mg$ the field $\vp$ is over-damped and it settles 
away from the origin at a distance 
\beq 
|\vp_0|\simeq \(c_H M^2 H^2 \over |\la|^2\)^{1/4}.
\eeq 
From the last equation, one sees that $\vp$ is displaced farther as the 
neutrino Yukawa coupling $\la$ is smaller. That is why $L_i$ in 
Eqt. (\ref{lh}) is usually chosen as the neutrino with the smallest 
Yukawa coupling, $L_1$ say. When $H \approx\mg$, the $A$-terms enter 
into play and the condensate begins to oscillate. In general, when taking 
into account thermal effects, the condensate begins to oscillate when the 
decreasing expansion rate reaches a certain value denoted $H_{\rm osc}$, 
determined when the thermal contributions are taken into account 
\cite{thermal}. At later times when $H\ll\mg$, 
the lepton number is essentially conserved. 
The evolution of the lepton number, $n_L$ defined as 
\beq
n_L={i\over 2}(\vp^*\dot{\vp}-\vp\dot{\vp^*}), 
\eeq
follows the equation
\beq
\label{nL}
\dot{n_L}+ 3 H n_L = \mbox{Im}\[\vp\, {\partial V(\vp) \over \partial \vp}
 \]
\eeq
The generated lepton asymmetry can be approximated by integrating the 
equation (\ref{nL}). This gives
\beq
n_L\approx {\mg \over 2 M}\mbox{Im}(a_m \vp^4)t
\eeq
In a matter dominated Universe, the expansion rate scales with time as 
$H=2/3 t$. Plugging this into the last equation, we get  
\beq
{n_L\over s}\approx {1\over 12}\({T_{\rm RH}\over H_{\rm osc}}\)
\(\mg \over \M \)
\({M\over \M}\) {\de_{\rm eff}\over |\la|^2}\,,
\eeq
where $\M \equiv M_{\rm Planck}/\sqrt{8 \pi} \simeq 2.4 \times
10^{18}$ GeV is the reduced Planck mass and we have dropped constants of 
\oo(1). The effective CP-violating 
parameter $\de_{\rm eff}$ is defined as 
\beq
\de_{\rm eff} \simeq \sin \(4\arg\vp+\arg a_m\)
\eeq
Now, specializing to the low scale see-saw models \cite{Borzumati,Arkani}, 
where Yukawa couplings come-out naturally suppressed as 
$\la \sim |Y^{\rm eff}|^2 \sim \mg / \M$, we get 
\beq
{n_L\over s}\approx {1\over 12}\({T_{\rm RH}\over H_{\rm osc}}\)\de_{\rm eff}
\eeq
Usually the effective CP-violating parameter is assumed to be maximal \ie 
$\de_{\rm eff}\simeq 1$. In our case there is no need to do so, since  
the reheating temperature can be as low as $\mg$ so TeV mass RHNs 
are produced thermally, while the gravitinos are not.
Typically, the condensate begins to oscillate at $H_{\rm osc}\gtrsim\mg$. 
In the extreme case when $T_{\rm RH}\simeq H_{\rm osc}\simeq \mg$, only a 
small amount of CP is sufficient to reproduce the observed value, namely
$
\de_{\rm eff}\simeq 10^{-9}-10^{-10}
$.
Up to now, we did not specify the transmission mechanism of SUSY breaking. We 
just assumed that some hidden sector will produce the 
soft breaking scalar masses and $A$-terms. In the gravity-mediated 
scenario the $A$-terms are known to be of the form $a \mg W + \mbox{h.c.}$. 
This means that $a_m=b_m \la$ and $a_H=b_H \la$, where $b_H,\,b_m\sim O(1)$. 
In this case, the resulting lepton asymmetry is given by
\beq
\label{ccp2}
{n_L\over s}\approx {1\over 12}\({T_{\rm RH}\over H_{\rm osc}}\)
\(\mg \over \M \) \de_{\rm eff}
\eeq
where now the CP violation parameter is defined as
\beq
\de_{\rm eff} \simeq \sin \(4\arg\vp+\arg a_m+ \arg\la\)
\eeq

For the typical values $H_{\rm osc}\sim\mg$, 
$T_{\rm RH}\sim 10^9\gev$ and $\de_{\rm eff}\sim O(1)$, we obtain the right 
value for the lepton asymmetry.

\section{Conclusions}
\label{sec:concl}
To conclude, motivated by the potential conflict between the gravitino 
overproduction bound and the high reheat temperature required to produce 
RHNs thermally, we investigated the baryogenesis through leptogenesis 
scenario in the presence of low scale RHNs. We have seen that, in such a 
scenario the Yukawa couplings of RHNs have to be suppressed, in order 
to give rise to acceptable light neutrino masses. This suppression
proved to be useful for many purposes, in particular in satisfying the 
out-of-equilibrium condition in the FY scenario. Due to this
suppression, however, the resulting CP was too small. We used two
different mechanisms to enhance the CP parameter: the degeneracy of
RHNs and soft $A$-terms. In the latter case, the necessary CP
violation may come entirely from the soft SUSY $A$-term. We also
considered leptogenesis via the $LH_u$ flat direction. We have seen
that for generic SUSY breaking scenarios, AD leptogenesis with low 
scale RHNs is possible, though with reheat temperatures higher than TeV.  

\section*{Acknowledgments}
I would like to thank B. Bajc, J. March-Russell, F. Vissani for useful 
discussions. Special thanks to G. Senjanovi\'c for careful reading of 
an earlier version of the manuscript and for valuable suggestions. I would 
like also to acknowledge the hospitality of the CERN theory division
where part of this work was done. This work was partially supported by 
the European RTN project ``Supersymmetry And The Early Universe''
under contract number HPRN-CT-2000-00152 and by the Italian INFN under 
the project ``Fisica Astroparticellare''.

\end{document}